\documentclass[twocolumn,secnumarabic,amssymb, nobibnotes,superscriptaddress, longbibliography, prb]{revtex4-1}

\usepackage{graphicx}
\usepackage{dcolumn}
\usepackage{bm}
\usepackage{amsmath}
\usepackage[detect-weight=true, detect-family=true]{siunitx}
\usepackage{xcolor}
\usepackage[normalem]{ulem}
\usepackage{hyperref}
\hypersetup{colorlinks=false}
\usepackage{subfiles}

\begin{document}

\preprint{AIP/123-QED}

\title{A graphene transmon operating at  \SI{1}{\tesla}}

\author{J. G. Kroll}
\affiliation{ 
	QuTech and Kavli Institute for Nanoscience, Delft University of Technology, Delft, 2600 GA, The Netherlands
}%
\author{W. Uilhoorn}%
\affiliation{ 
	QuTech and Kavli Institute for Nanoscience, Delft University of Technology, Delft, 2600 GA, The Netherlands
}%
\author{K. L. van der Enden}
\affiliation{ 
	QuTech and Kavli Institute for Nanoscience, Delft University of Technology, Delft, 2600 GA, The Netherlands
}%

\author{D. de Jong}
\affiliation{ 
	QuTech and Kavli Institute for Nanoscience, Delft University of Technology, Delft, 2600 GA, The Netherlands
}%

\author{K. Watanabe}
\affiliation{ 
Advanced Materials Laboratory, National Institute for Materials Science, 1-1 Namiki, Tsukuba, 305-0044, Japan
}%

\author{T. Taniguchi}
\affiliation{ 
Advanced Materials Laboratory, National Institute for Materials Science, 1-1 Namiki, Tsukuba, 305-0044, Japan
}%

\author{S. Goswami}
\affiliation{ 
	QuTech and Kavli Institute for Nanoscience, Delft University of Technology, Delft, 2600 GA, The Netherlands
}%

\author{M. C. Cassidy}
\affiliation{ 
	Microsoft Station Q Sydney, Sydney, NSW 2006, Australia
}%

\author{L. P. Kouwenhoven}
 \email{Leo.Kouwenhoven@Microsoft.com}
\affiliation{ 
	QuTech and Kavli Institute for Nanoscience, Delft University of Technology, Delft, 2600 GA, The Netherlands
}%
\affiliation{ 
	Microsoft Station Q Delft, Delft, 2600 GA, The Netherlands
}%

\date{\today}

\pacs{Valid PACS appear here}
\maketitle
\noindent

\textbf{A superconducting transmon qubit \cite{Koch2007} resilient to strong magnetic fields is an important component for proposed topological  \cite{Hassler2011,Hyart2013, Aasen} and hybrid quantum computing (QC) schemes \cite{Kubo2011, Ranjan2013}. Transmon qubits consist of a Josephson junction (JJ) shunted by a large capacitance, coupled to a high quality factor superconducting resonator. In conventional transmon devices, the JJ is made from an Al/AlO$_x$/Al tunnel junction \cite{Koch2007} which ceases operation above the critical magnetic field of Al, $\sim$\SI{10}{\milli \tesla}. Alternative junction technologies are therefore required to push the operation of these qubits into strong magnetic fields. Graphene JJs are one such candidate due to their high quality, ballistic transport and electrically tunable critical current densities \cite{Calado2015, Allen2015,Chtchelkatchev2007}. Importantly the monolayer structure of graphene protects the JJ from orbital interference effects that would otherwise inhibit operation at high magnetic field. Here we report the integration of ballistic graphene JJs into microwave frequency superconducting circuits to create the first graphene transmons. The electric tunability allows the characteristic band dispersion of graphene to be resolved via dispersive microwave spectroscopy. We demonstrate that the device is insensitive to the applied field and perform energy level spectroscopy of the transmon at \SI{1}{\tesla}, more than an order of magnitude higher than previous studies \cite{Ku2016,Luthi2017}.}

Despite their antagonistic relationship, superconductivity and a magnetic field are key ingredients in current hybrid and topological QC proposals \cite{Lutchyn2010, Kubo2011}. This combination proves challenging for practical implementations as a magnetic field causes undesirable effects in superconductors, such as reduction of the superconducting gap, increased quasiparticle generation \cite{VanWoerkom2015} and the formation of Abrikosov vortices that cause resistive losses in a microwave field. In addition to disrupting the superconductivity, magnetic flux penetrating the JJ produces electron interference effects that reduce the Josephson energy $E_\text{J}$ and strongly suppress the transmon energy spectrum. If the transmon is to be used for fast quantum gates, fast charge-parity detection and long range quantum state transfer in QC schemes \cite{Kurizki, Riste2012,Hyart2013} we are compelled to consider alternatives to conventional Al based JJs. Proximitised semiconducting nanowires, acting as gate-tuneable superconductor-normal-superconductor JJs \cite{Doh2005} have been used successfully in a variety of microwave frequency superconducting circuits, allowing for studies of Andreev bound states \cite{VanWoerkom2017, Hays2017}, electrically tuneable transmon qubits \cite{delange2015,larsen2015} and transmons that exhibit substantial field compatibility \cite{Luthi2017}. Graphene JJs are an attractive alternative as they exhibit ballistic transport, high critical currents and the atomic thickness of the graphene junction greatly reduces flux penetration, protecting $E_\text{J}$ in high parallel fields. When combined with geometric techniques to protect the superconducting film, such as critical field enhancement \cite{Stan2004} and lithographically defined vortex pinning sites \cite{Bothner2011}, the transmon can be protected at magnetic fields relevant to these proposals, which in some cases exceeds \SI{1}{\tesla}.

\begin{figure}
	\includegraphics[width=0.5\textwidth]{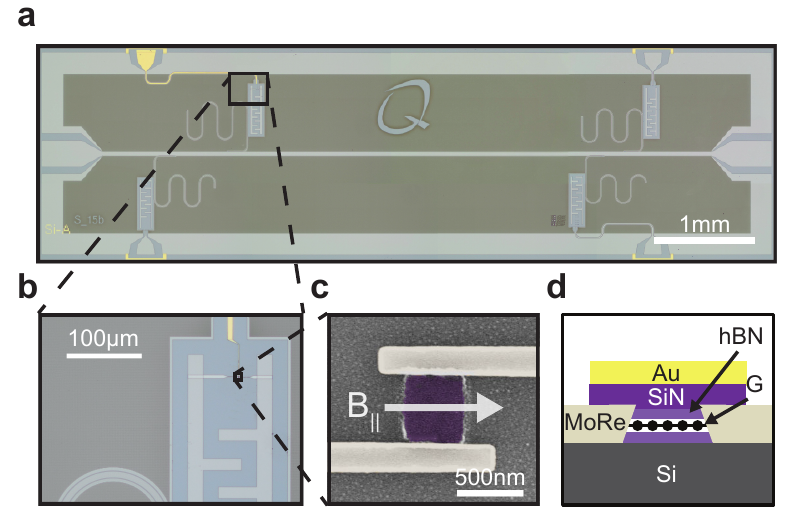}
	\caption{(a) Optical image showing multiple CPW resonators frequency multiplexed to a common feedline (device B). (b) Zoomed optical image of the capacitor plates that shunt the Josephson junction, with the gate, junction and contacts visible. (c) SEM micrograph of a contacted boron nitride-graphene-boron nitride stack before fabrication of the gate. A magnetic field $B_{||}$ can be applied parallel to the film along the length of the junction contacts using a 3-axis vector magnet. (d) Cross sectional diagram showing the fully contacted and gated stack.}
	\label{deviceoverview}
\end{figure}
Fig.~\ref{deviceoverview}a shows an optical microscope image of a typical graphene transmon device. It consists of four $\lambda$/4 coplanar waveguide (CPW) resonators multiplexed to a common feedline. Each resonator is capacitively coupled to a graphene transmon, with the graphene JJ being shunted by capacitor plates that provide a charging energy $E_\text{C} \simeq $ \SI{360}{\mega \hertz}. The resonators and capacitor plates are fabricated from \SI{20}{\nano \meter} NbTiN due to its enhanced critical magnetic field \cite{Stan2004}, and we pattern the resonators with a lattice of artificial pinning sites to protect the resonator from resistive losses due to Abrikosov vortices \cite{Bothner2011}. The van der Waals pickup method  is used to fabricate the graphene JJ by encapsulating monolayer graphene between two hexagonal boron nitride (hBN) flakes and depositing it between pre-fabricated capacitors plates (Fig.~\ref{deviceoverview}b) \cite{Calado2015}. Contact to the hBN/G/hBN stack is made by dry etching and sputtering MoRe. In this work, we present results from two graphene JJ transmon devices, with slightly different fabrication techniques. Device A uses a Ti/Au gate stack deposited directly on the hBN, before the junction is shaped via dry etching. Device B is shaped (Fig.~\ref{deviceoverview}c) before a Ti/Au gate stack with a SiN$_x$ interlayer is deposited (Fig.~\ref{deviceoverview}d). The sample is mounted inside a light-tight copper box and thermally anchored to a dilution refrigerator with a base temperature of \SI{15}{\milli \kelvin}. An external magnetic field is applied to the sample using a 3-axis vector magnet. The complex microwave transmission $S_{21}$ is measured using standard heterodyne demodulation techniques. A second tone can be injected to perform two-tone spectroscopy of the system. Detailed fabrication and measurement notes can be found in the Supplementary Information.

\begin{figure}
	\includegraphics[width=0.5\textwidth]{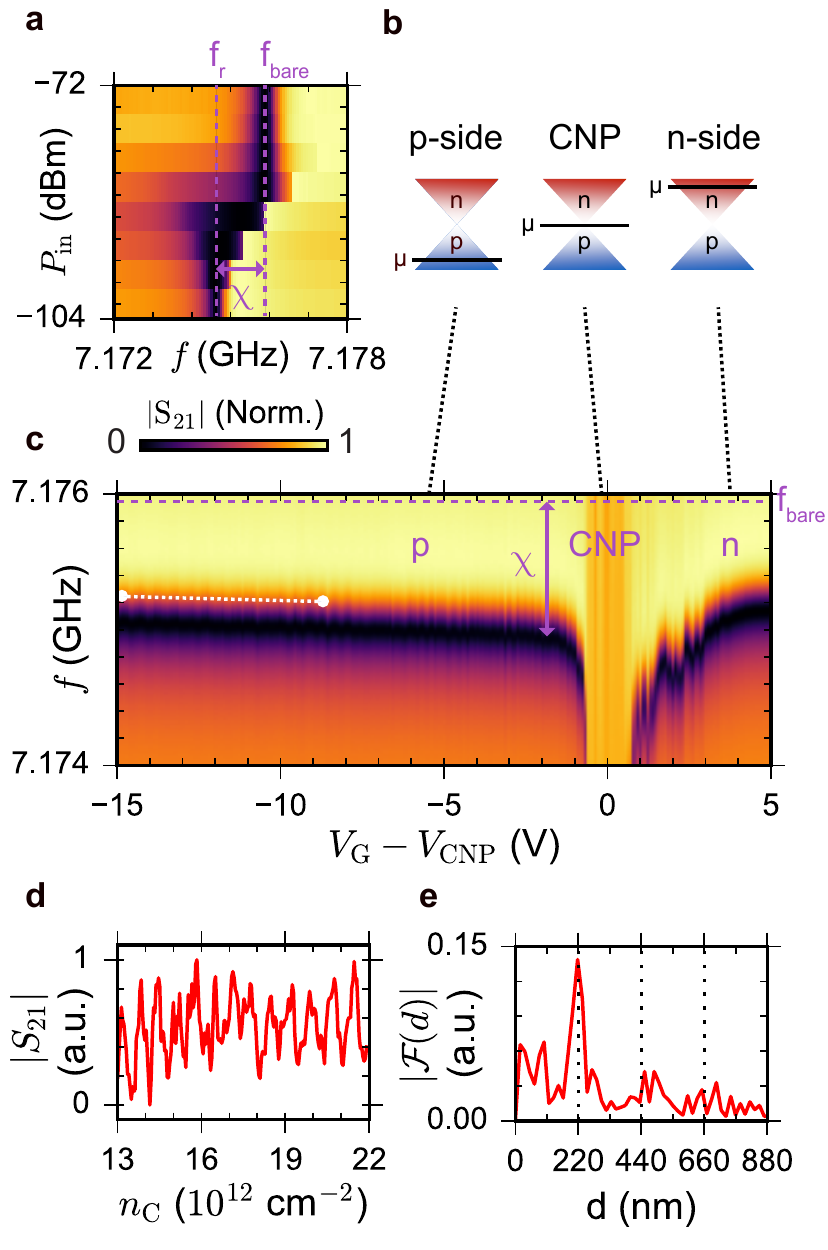}
	\caption{(a) $|S_{21}|$ (Norm.) as a function of input frequency $f$ and input power $P_{\text{in}}$. At single photon occupancy the resonator experiences a frequency shift $\chi$ due to repulsion from an energy level above the resonator (device A). (b) Diagram of the Dirac cone band structure of graphene. Changing $V_{\text{G}}$ to tune $\mu$ allows the dominant charge carriers to be varied between hole, charge neutral and electron-like regimes. (c) At single photon occupancy, $|S_{21}|$ (Norm.) is measured as $f$ and $V_{\text{G}}$ are varied, with the voltage at CNP ($V_{\text{CNP}}$) subtracted. In the p-regime, $\chi$ oscillates as $V_{\text{G}}$ is varied. We extract the charge carrier density $n_\text{c}$ (d) from the white linecut to generate a Fourier transform (e) that is consistent with Fabry-Perot oscillations in a cavity of $d = \SI{220}{\nano \meter}$.}
	\label{devicea}
\end{figure}

We begin by performing spectroscopy of the resonator in device A as a function of the input power $P_\text{in}$ (Fig.~\ref{devicea}a). Varying the resonator's photon occupation from $\langle n_{\text{ph}} \rangle \simeq 1000$ to $\langle n_{\text{ph}} \rangle = 1$ we observe a shift $\chi = f_\text{r} - f_\text{bare}$ in the resonator frequency $f_\text{r}$ from the high power value $f_\text{bare}$. This occurs due to a Jaynes-Cummings type interaction between the harmonic readout resonator and the anharmonic transmon spectrum, with the anharmonicity provided by the Josephson junction \cite{Reed}. The magnitude of the shift $\chi = g^2/\Delta$ depends on the transmon-resonator coupling $g$, and the difference $\Delta = f_\text{r}-f_{\text{t}}$ between ${f_\text{r}}$ and the ground state to first excited state transition frequency ${f_\text{t}} = E_\text{t}/h \simeq \sqrt{8 E_\text{J} E_\text{C}}/h$, allowing us to infer $E_\text{J}$ from $\chi$ \cite{Koch2007}. Studying $\chi$ as a function of $V_{\text{G}}$ reveals the characteristic band dispersion of graphene (Fig.~\ref{devicea}b). At negative $V_{\text{G}} - V_{\text{CNP}}$, the chemical potential $\mu$ is below the charge neutrality point (CNP) and the graphene is in the p-regime where holes are the dominant charge carrier. Deep into the p-regime, the high carrier density ($n_\text{C}$) gives a large $E_\text{J}$, placing $f_\text{t}$ above the resonator and giving $\chi$ a small negative value (Fig.~\ref{devicea}c). As $V_{\text{G}}$ approaches the CNP, the Dirac dispersion minimises the density of states reducing $E_\text{J}$ to a minimum, causing $\chi$ to diverge. As $V_{\text{G}}$ is increased past the CNP, electrons become the dominant charge carrier and $E_\text{J}$ increases to a maximum, as expected from removal of the n-p-n junction formed by the contacts \cite{Calado2015}. The p-regime also experiences periodic fluctuations in $E_\text{J}$ as a function of  $V_{\text{G}}$ due to electron interference effects in a Fabry-Perot cavity formed by n-p interfaces at the MoRe contacts \cite{Calado2015}. Extracting a line trace (white line Fig.~\ref{devicea}c) to study the modulation in $|S_{\text{21}}|$ with $n_\text{C}$ (Fig.~\ref{devicea}d), and performing a Fourier transform (Fig.~\ref{devicea}e) gives a cavity length of \SI{220}{\nano \meter} in agreement with the device dimensions. The observation of a Dirac dispersion relation in combination with coherent electron interference effects confirm the successful integration of ballistic graphene JJs into a superconducting circuit. 

 \begin{figure}
	\includegraphics[width=0.5\textwidth]{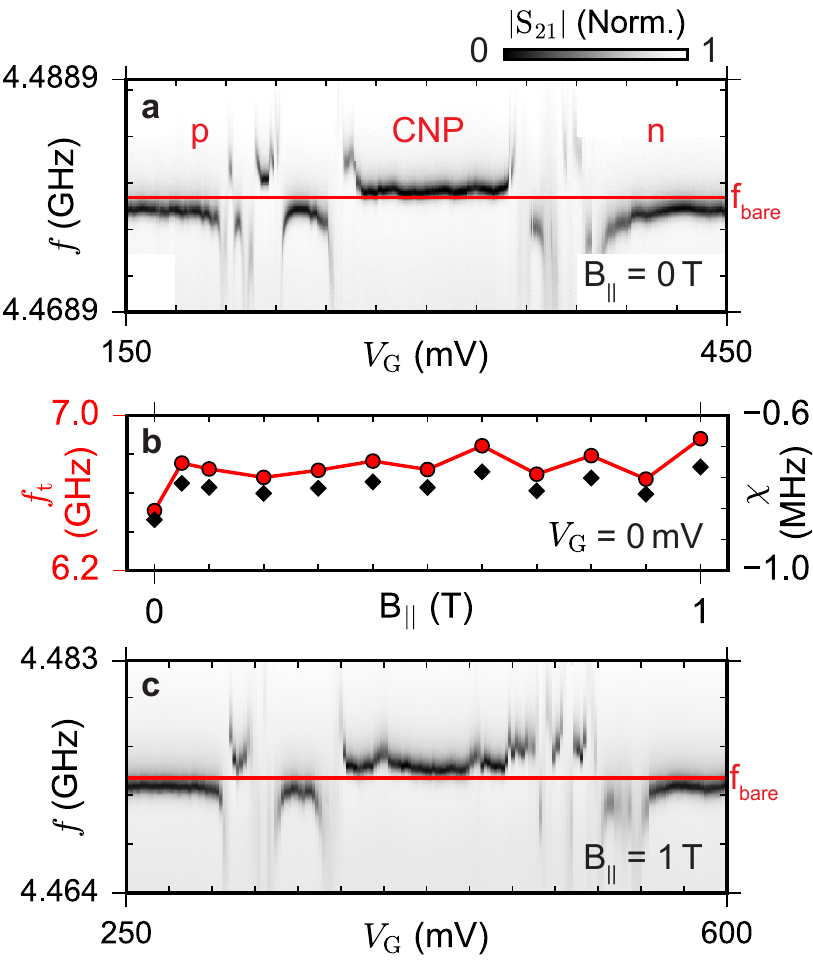}
	\caption{(a) At $B_{||} = \SI{0}{\tesla}$, $|S_{21}|$ (Norm.) versus $f$ and $V_\text{G}$ shows the band dispersion of graphene with additional fluctuations we attribute to UCF. (b) $f_\text{t}$ (red circles) extracted from $\chi$ (black diamonds) versus $B_{||}$ at $V_\text{G} = \SI{0}{\volt}$, showing $f_\text{t}$ is not significantly affected. (c) Repeating (a) at $B_{||} = \SI{1}{\tesla}$ confirms the graphene JJ behaves equivalently to $B_{||} = \SI{0}{\tesla}$. The variation observable in (b) and shift in $V_\text{G}$ between (a) and (c) we attribute to slow gate drift.}
	\label{fielddependence}
\end{figure}

\begin{figure}
\includegraphics[width=0.5\textwidth]{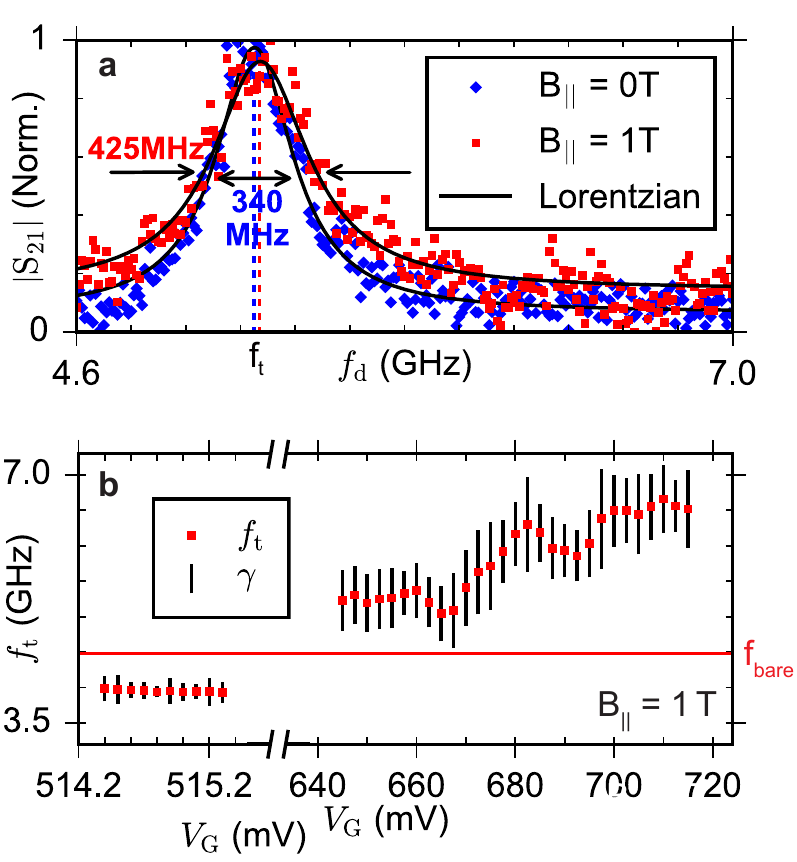}
\caption{(a) Normalised $|S_{21}|$ at $f_\text{r}$ as $f_\text{d}$ is varied can be fitted to extract $f_\text{t}$ and $\gamma$ at $V_\text{{G}}$ = \SI{0}{\volt}. At $B_{||} = $\SI{1}{\tesla}, $\gamma$ shows a 25\% increase compared to $B_{||} = $ \SI{0}{\tesla}. (b) At $B_{||} = $\SI{1}{\tesla}, $f_\text{t}$ and $\gamma$ are extracted as $V_\text{G}$ is varied, demonstrating $f_\text{t}$ can be swept over a wide frequency range.}
\label{spectroscopy}
\end{figure}

In device B we observe additional coherent electronic interference effects in the form of universal conductance fluctuations (UCF). As we move from the p to the CNP regime, $\chi$ is seen to diverge repeatedly as $f_\text{t}$ anti-crosses multiple times with $f_\text{r}$ (Fig.~\ref{fielddependence}a). This behaviour is repeated moving from the CNP to the n-regime, where $E_\text{J}$ is again maximised. We demonstrate the field compatibility of the junction by applying a magnetic field $B_{||}$ along the length of the junction contacts, parallel to the plane of the film, using the resonator as a sensor for field alignment (see Supplementary Information for alignment procedure details). Monitoring $\chi$ as $B_{||}$ is varied between 0 and \SI{1}{\tesla}  (Fig.~\ref{fielddependence}b) and calculating $f_\text{t}$ (using $g = \SI{43}{\mega \hertz}$, extracted from measurements in Fig.~\ref{spectroscopy}), demonstrates that $\chi$ and thus $E_\text{J}$ are not significantly affected by the applied $B_{||}$. The small amount of variation observed is attributed to charge noise induced gate drift which was observed throughout the duration of the experiment. Studying $\chi$ as a function of $V_{\text{G}}$ at $B_{||}$ = \SI{1}{\tesla} (Fig.~\ref{fielddependence}c) again reveals the characteristic Dirac dispersion as seen in Fig.~\ref{fielddependence}a, with modified UCF. The insensitivity of $f_\text{t}$ to applied field and similarity of device operation at $B_{||} = $ \SI{0}{\tesla} and \SI{1}{\tesla} confirm the field resilience of both the graphene JJ and superconducting circuit.

In order to better understand the microwave excitation spectra of our system we proceed to measure it directly via two-tone spectroscopy. The readout tone is set to $f_\text{r}$ whilst a second tone $f_\text{d}$ is used to drive the circuit. Excitation of the system results in a state dependent shift of the resonator frequency \cite{Koch2007}, and is detected by measuring the change in the complex transmission $S_{\text{21}}$ at $f_\text{r}$. At $V_{\text{G}}$ = \SI{0}{\volt}, two-tone spectroscopy at $B_{||}$ = 0 and \SI{1}{\tesla} (Fig.~\ref{spectroscopy}a) can be fitted with a Lorentzian to extract the transmon transition $f_\text{t}$ and transition linewidth $\gamma$. At $B_{||}$ = \SI{1}{\tesla}, $f_\text{t}$ and thus $E_\text{J}$ differ only slightly with $\gamma$ increasing slightly from \SI{350}{\mega \hertz} to \SI{425}{\mega \hertz}. The transmon resonator coupling $g = \sqrt{\chi \Delta} =$ \SI{43}{\mega \hertz} is extracted from the observed dispersive shift $\chi$ and detuning $\Delta$, and used in the calculation of $f_\text{t}$ in Fig.~\ref{fielddependence}. We attribute the change in $f_\text{t}$ from Fig.~\ref{fielddependence}b and the large $\gamma$ to the dielectric induced charge noise mentioned previously. An estimate of $E_{\text{J}} = \SI{40.2}{\micro \electronvolt}$ can be provided using the relation $E_\text{t} = h f_\text{t} \simeq \sqrt{8E_{\text{J}}E_{\text{C}}}$. Additional measurement of the higher order two-photon $f_{02}$ transition would allow for exact measurements of $E_{\text{J}}$ and $E_{\text{C}}$ via diagonalisation of the Hamiltonian, enabling investigations into mesoscopic effects of interest in graphene JJs \cite{Nanda2017,Kringhøj2018}. Performing two-tone spectroscopy whilst tuning $V_{\text{G}}$ reveals a gate-tunable energy level that is visible on the p-side and n-side of the CNP, reiterating the electron-hole symmetry expected in graphene. The transition and thus $E_\text{J}$ can be varied over a wide frequency range, satisfying a key requirement for implementation into topological QC proposals \cite{Hyart2013} (Fig.~\ref{spectroscopy}b). The large linewidths suggest that although measurements of relaxation and coherence lifetimes  ($T_1, T_2^*$) may be possible, it is experimentally challenging.

In conclusion, we have demonstrated the first integration of a graphene JJ into a superconducting circuit to make a graphene based transmon. Additionally, we have achieved operation at $B_{||}$ = \SI{1}{\tesla}, a magnetic field more than an order of magnitude higher than any previous transmon \cite{Ku2016,Luthi2017}. While the broad linewidths prevented the demonstration of coherent qubit control, these results establish graphene based microwave circuits as a promising tool for topological and hybrid QC schemes, and for probing mesoscopic phenomena of interest at high magnetic fields.

\begin{acknowledgments}
We thank D.J. van Woerkom for fabrication assistance, M.W.A. de Moor and A. Proutski for helpful discussion and L. DiCarlo, C. Dickel and F. Luthi for experimental advice and software support. This work has been supported by the European Research Council (ERC), The Dutch Organisation for Scientific Research (NWO) and Microsoft Corporation Station Q.
\end{acknowledgments}

\section*{Author contributions}
K.W. and T.T. grew the hBN crystals, J.G.K. and W.U. fabricated the devices, J.G.K., K.L.v.d.E. and D.d.J performed the measurements and J.G.K. and K.L.v.d.E. analysed the measurements. The manuscript was prepared by J.G.K. with K.L.v.d.E., S.G., M.C. and L. P. K. providing input. S.G., M.C. and L.P.K. supervised the project.

\clearpage
\setcounter{figure}{0}
\renewcommand{\thefigure}{S\arabic{figure}}
\onecolumngrid
\section{\large{Supplementary Information}}
\vspace{10mm}

\newcommand{\beginsupplement}{%
	\setcounter{table}{0}
	\renewcommand{\thetable}{S\arabic{table}}%
	\setcounter{figure}{0}
	\renewcommand{\thefigure}{S\arabic{figure}}%
}

\beginsupplement

\maketitle
\section{Device fabrication}
\begin{center}
	\begin{figure}[htb!]
		\includegraphics[width=0.6\textwidth]{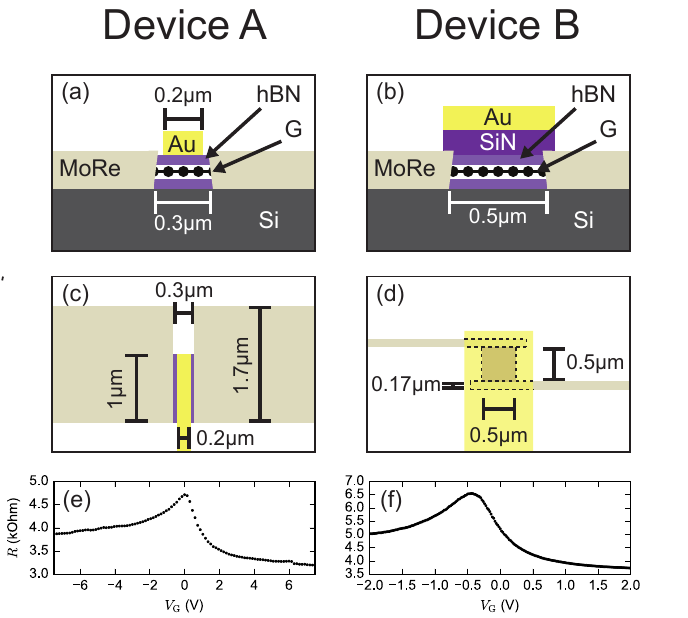}
		\caption{(a) Cross sectional diagram showing the geometry of the junction in device A. It is \SI{300}{\nano \meter} in length, with a gate designed to be  \SI{200}{\nano \meter} in width. (b) Cross sectional diagram of device B, with the SiN$_x$/Ti/Au gate giving full coverage of the \SI{500}{\nano \meter} long junction. (c) The junction in device A is \SI{1000}{\nano \meter} wide and contacted by \SI{1700}{\nano \meter} wide MoRe contacts. The \SI{200}{\nano \meter} wide gate can be seen to cover most, but not all of the graphene stack. (d) Device B has a 500 x \SI{500}{\nano \meter \squared} junction that is contacted by thin \SI{170}{\nano \meter} MoRe leads to prevent vortices from forming near the junction. The SiN$_x$/Ti/Au gate stack is sputtered and designed to give full coverage of the graphene junction. (e) Measurement of the two point resistance $R$ of the contacts and junction for device A at room temperature as the gate voltage $V_\text{G}$ is varied. The charge neutrality point can be observed at $V_\text{G} = \SI{0}{\volt}$. (f) $R$ measurements for device B at room temperature again showing the charge neutrality point, this time offset slightly to  $V_\text{G} \simeq \SI{-0.5}{\volt}$. Upon cooling to $T = \SI{15}{\milli \kelvin}$, the charge neutrality points were observed to shift in both devices. }
		\label{FBdev}
	\end{figure}
\end{center}
Two separate devices, device A and device B were used throughout the manuscript. For device fabrication, the type II superconductors NbTiN and MoRe were chosen for their high upper critical fields ($B_{c_{2}} >  \SI{8}{\tesla}$) and their compatibility with microwave frequency devices \cite{Megrant2012,Singh2014}. Resistive losses from Abrikosov vortices at microwave frequencies are mitigated by expelling the vortices via geometric constriction \cite{Stan2004,Samkharadze2016} and using artificial pinning sites to trap the vortices that cannot be excluded \cite{Bothner2011}. To fabricate the devices \SI{20}{\nano \meter} of NbTiN is sputtered onto intrinsic Si wafers in an Ar/N atmosphere. The resonators, feedline and transmon are reactive ion etched in an SF$_6$/O$_2$ atmosphere. In this etching step, an array of artificial pinning sites is also defined. Monolayer graphene is encapsulated between two hBN flakes ($t$ $\simeq$ \SI{15}{\nano \meter} each), then deposited between pre-fabricated capacitors using a PMMA based van der Waals pickup method \cite{Calado2015}. Contact to the graphene stack is made by etching in a CHF$_3$/O$_2$ environment, followed by sputtering MoRe ($t$ = \SI{80}{\nano \meter}). Device A was contacted to give a junction length of \SI{300}{\nano \meter} (Fig~\ref{FBdev}a,c). A Ti/Au top gate is then sputtered on top of the stack. The device is then shaped in a CHF$_3$/O$_2$ plasma to be 1000 x \SI{300}{\nano \meter \squared} in size. Device B was contacted to provide a junction length of \SI{500}{\nano \meter} (Fig~\ref{FBdev}b,d). The long thin leads were geometrically restricted in two dimensions, making it less favourable for vortices to form, protecting the superconductivity of the contacts  proximitising the junction. The junction is then shaped in a CHF$_3$/O$_2$ plasma to be 500 x \SI{500}{\nano \meter \squared}. A SiN$_x$/Ti/Au top gate stack is then sputtered to give full junction coverage, giving greater control of $\mu$ in the junction. A probe station was used to perform two probe resistance $R$ measurements of the graphene junction and contact resistances at room temperature. Device A (Fig~\ref{FBdev}e) and device B (Fig~\ref{FBdev}f) both show charge neutrality points in the $R$ vs $V_\text{G}$ dependences, consistent with graphene junctions. 

\section{Experimental circuit}
\begin{center}
	\begin{figure}[h]
		\includegraphics[width=0.9\textwidth]{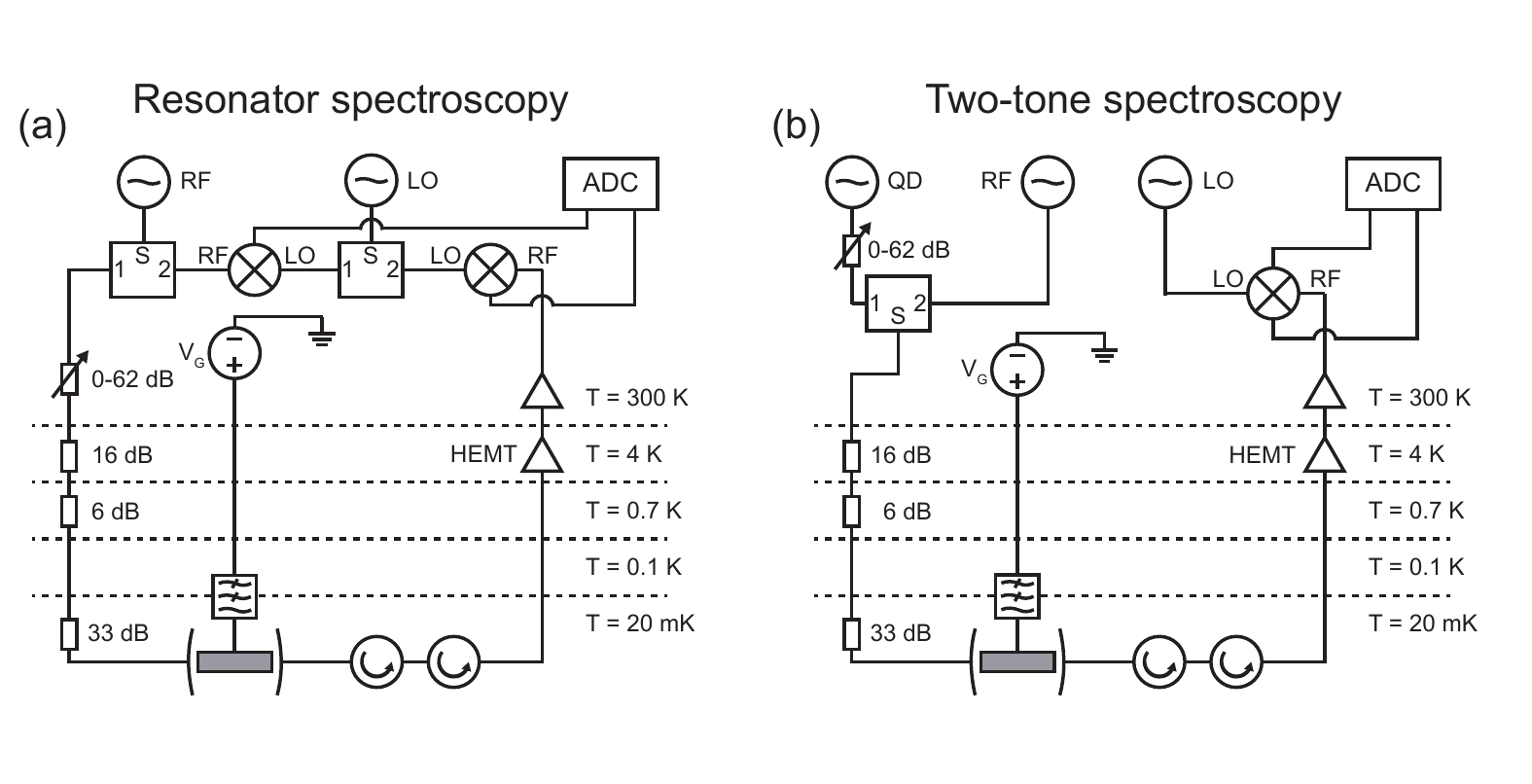}
		\caption{Diagram showing the DC and microwave measurement circuits.}
		\label{measurementcircuit}
	\end{figure}
\end{center}

All measurements were performed in a dilution refrigerator with a base temperature of \SI{15}{\milli \kelvin}. The samples were enclosed in a light tight copper box, and thermally anchored to the mixing chamber. The two measurement configurations used are depicted in Fig. \ref{measurementcircuit}. Two coaxial lines and one DC line were used to control the sample. The sample was connected to the DC voltage source by a line that was thermally anchored at each stage and heavily filtered at the mixing chamber by low RC, $\pi$ and copper powder filters. The line used to drive the feedline input was heavily attenuated to reduce noise and thermal excitation of the cavity, allowing the single photon cavity occupancy to be reached. The output line of the feedline was connected to an isolator (Quinstar QCI-080090XM00) and circulator (Quinstar QCY-060400CM00) in series to shield the sample from thermal radiation from the HEMT amplifier (Low Noise Factory LNF-LNC4-8\_C) on the 4K stage. 

Resonator spectroscopy of device A was performed using circuit (a) to measure the amplitude and phase response of the complex transmission $S_{21}$ as the frequency was varied. Resonator and two-tone spectroscopy of device B was performed using circuit (b), with a splitter used to combine the readout and excitation tones. This allows the complex $S_{21}$ to be measured, but only at fixed resonator readout frequency otherwise only $|S_{21}|$ can be recorded.

\section{Readout circuit resonance}

\begin{center}
	\begin{figure}[h]
		\includegraphics[width=0.95\textwidth]{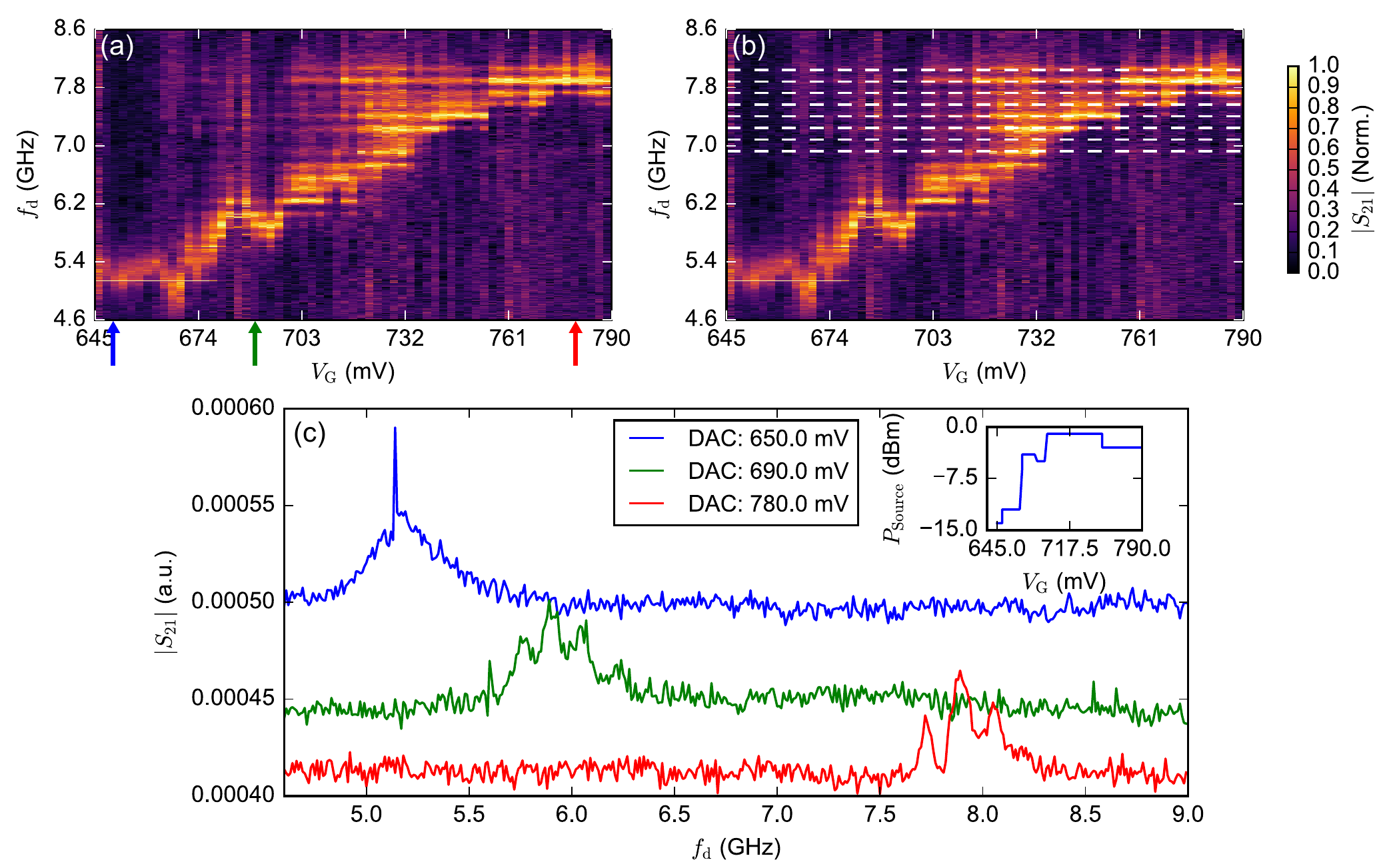}
		\caption{(a) Normalised heat map of $S_{21}$ during two-tone spectroscopy as the gate voltage $V_\text{G}$ and drive tone $f_\text{d}$ is varied. (b) The response of $f_\text{t}$ is modulated by background resonances due to a standing wave in the readout circuit, as evidenced by the white lines. (c) Line traces showing $|S_{21}|$ at several $V_\text{G}$ values (corresponding to linecuts at coloured arrows in (a)). The sharp resonance at \SI{5.143}{\giga \hertz} is due to interference effects from an additional $\lambda /4$ resonator multiplexed to the common feedline. The inset demonstrates that above $\sim$ \SI{670}{\milli \volt} the higher power required to drive the transition excites additional modes in the circuit, giving the modulation in response as seen in (a). Above \SI{720}{\milli \volt} the resonances become so extreme that it is difficult to analyse the data reliably, so they are excluded from the main text.}
		\label{savgol}
	\end{figure}
\end{center}

Normalised heatmap of $|S_{21}|$ during two-tone spectroscopy of the transmon energy level $f_\text{t}$ as the gate voltage $V_\text{G}$ is varied (Fig \ref{savgol}a). A sharp peak (dip) at \SI{5.143}{\giga \hertz} (\SI{5.640}{\giga \hertz}) is visible due to resonant driving of additional $\lambda/4$ resonators multiplexed to the same feedline. The drive is provided to the transmon indirectly, through the coupled resonator. As the detuning $\Delta = f_\text{t} - f_\text{r}$ is increased, the amount of drive power required to excite the transition at $f_\text{t}$ also increases due to the filtering effect of the resonator (Fig \ref{savgol}c inset) and low relaxation time $T_1$. Above \SI{5.7}{\giga \hertz} the drive power becomes so high that an additional resonant mode in the circuit is excited, modulating the two-tone response of $S_{21}$. We attribute these oscillations to a standing wave ($\Delta f = $ \SI{160}{\mega \hertz}, $\lambda$ = \SI{1.25}{\meter}) in the coaxial cables caused by an impedance mismatch at the device (Fig \ref{savgol}b). The resonance was only observable whilst near $f_\text{t}$ and at very high power, making a simple background subtraction difficult. Above $\sim$\SI{720}{\milli \volt} the very high power required to drive the transition causes many resonances, complicating the analysis. Due to this, data above this point was excluded from the analysis in the main text.

\section{Magnetic field alignment}
\label{fieldalignment}
The superconducting coplanar waveguide resonators are fabricated from a type II superconductor NbTiN, which allows flux to penetrate it in the form of Abrikosov vortices. These vortices experience a Lorentz force under microwave irradiation, causing them to oscillate in the superconductor and cause losses and fluctuations in the resonators inductance and frequency. Although the resonators have been patterned with etched pinning sites to withstand a certain vortex density (\SI{1.8}{\micro\meter}$^{-2}$), even a small misalignment at \SI{1}{\tesla} can exceed this, creating more vortices than can be successfully pinned. This results in degradation of the resonator performance. To compensate for this, a 3-axis vector magnet was used to align the applied field parallel to the film of the superconducting resonators.
\begin{center}
	\begin{figure}[h]
		\includegraphics[width=1\textwidth]{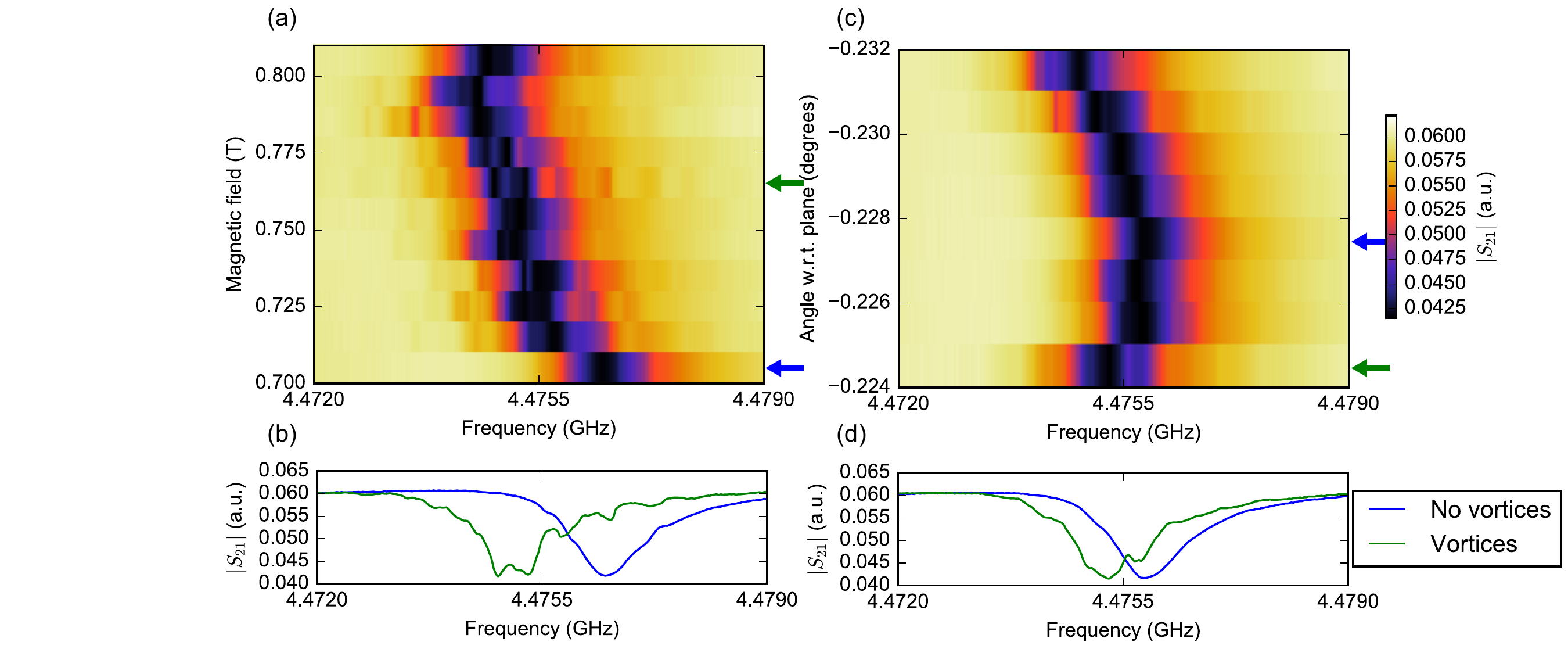}
		\caption{(a) Measuring the feedline transmission $S_{21}$ as a function of probe frequency $f$ as the magnetic field $B$ is increased from \SI{0.7}{\tesla} to \SI{0.8}{\tesla}. (b) As $B$ increases the resonator that was previously stable (blue arrow/linecut) is affected by unpinned Abrikosov vortices (green arrow/linecut) giving fluctuations in the resonator frequency on the timescale of the experiment. (c) Varying the angle of the magnetic field with respect to the plane of the superconducting film and measuring the response of the resonator allows the field to be realigned. (d) Although the sweep in magnetic field is hysteretic, the angle is varied until the resonator that was previously affected by vortices (green arrow/linecut) reaches a maximum $f_\text{r}$ and the fluctautions cease (blue arrow/linecut).}
		\label{alignment}
	\end{figure}
\end{center}
During mounting of the sample, the orientation of the device is estimated so that the field can be applied approximately parallel to the film. Nucleation of vortices in the film results in the frequency of the resonator and the quality factor $Q_\text{i}$ reducing. An example alignment procedure when moving from 0.7 to \SI{0.8}{\tesla} can be observed in Fig~\ref{alignment}. In (a) at \SI{0.7}{\tesla} the resonator is at a maximum in frequency with a well-defined. When increasing the field, misalignment causes vortex penetration that reduces the resonant frequency $f_\text{r}$ and causes it to fluctuate on the time scale of the experiment, giving significant modulations in the line shape. At \SI{0.8}{\tesla} the angle with respect to the plane of the superconducting film is varied until $f_\text{r}$ reaches  maximum and the line shape fluctuations cease. Using this method, we are able to retain stable resonances and single photon internal quality factors exceeding 100,000 at $B_{||} = \SI{6}{\tesla}$.

\section{Lead orientation}

\begin{center}
	\begin{figure}[h]
		\includegraphics[width=0.6\textwidth]{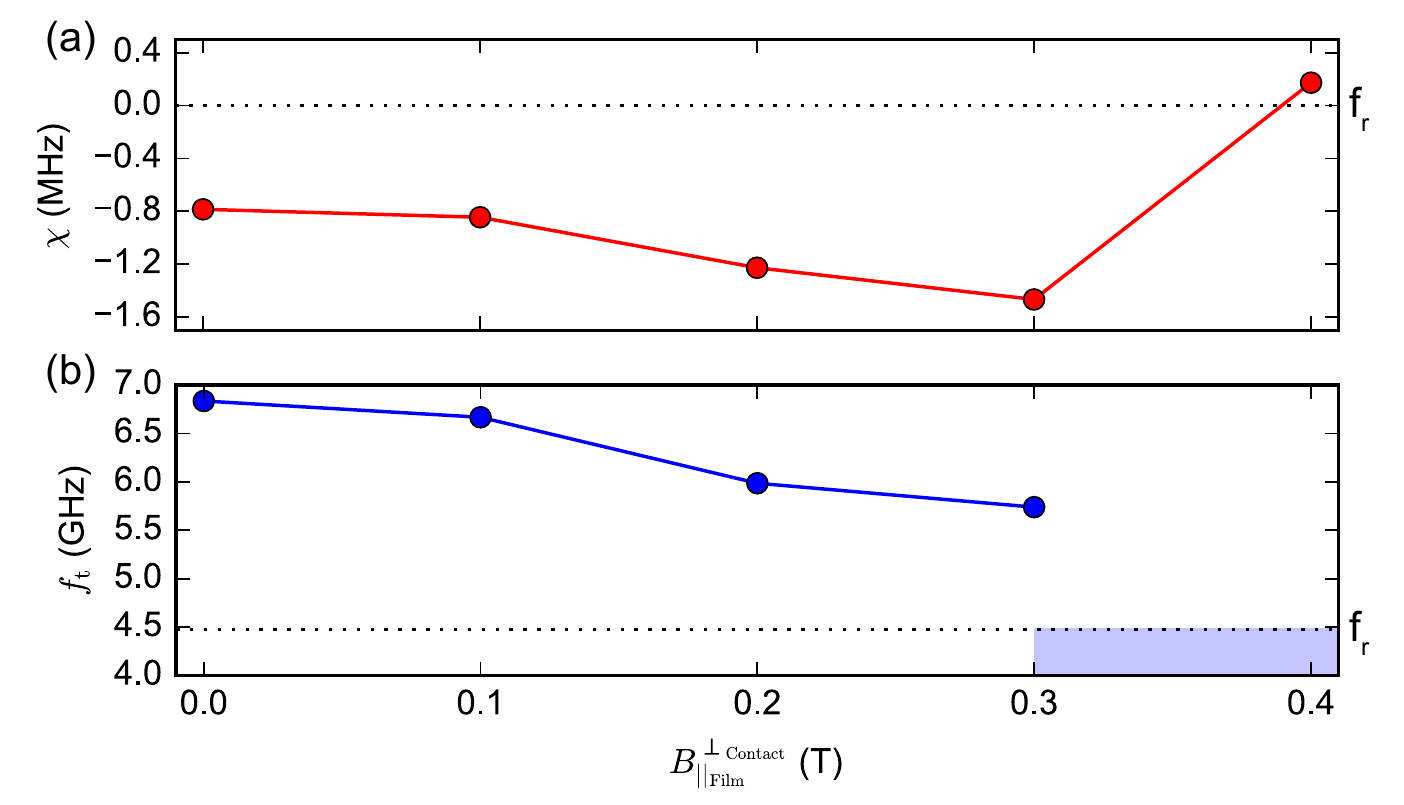}
		\caption{(a) Dependence of dispersive shift $\chi$ with respect to field $B^{\perp_{\text{Contact}}}_{||_{\text{Film}}}$ applied perpendicular to the length of the junction contacts, but parallel to the film. (b) Extracting $f_\text{t}$ from $\chi$ shows that as $B^{\perp_{\text{Contact}}}_{||_{\text{Film}}}$ is applied the $E_\text{J}$ reduces significantly. As $f_\text{t}$ approaches $f_\text{r}$, $\chi$ increases until $f_\text{t}$ passes through $f_\text{r}$ resulting in the sign change of $\chi$. Above $B^{\perp_{\text{Contact}}}_{||_{\text{Film}}}$ = \SI{0.4}{\tesla} the dispersive regime is no longer valid, meaning $f_\text{t}$ cannot be accurately estimated. These results imply that the $E_\text{J}$ of the junction is only protected if $B$ is applied along the length of the junction contacts.}
		\label{leadalignment}
	\end{figure}
\end{center}

During measurement it was found that the orientation of the field with respect to the leads was of key importance. Initially the field was applied along the main axis of the 6-1-\SI{1}{\tesla} vector magnet; parallel to the film and perpendicular to the junction contacts ($B^{\perp_{\text{Contact}}}_{||_{\text{Film}}}$). Despite careful alignment as described in Sec. \ref{fieldalignment} the Josephson energy ($E_\text{J}$) was found to significantly reduce at only $\sim $ \SI{400}{\milli \tesla}. In Fig \ref{leadalignment}b as $B^{\perp_{\text{Contact}}}_{||_{\text{Film}}}$) is increased, $E_\text{J}$ reduces, causing the transmon transition to also reduce (as $f_\text{t} \simeq \sqrt{8 E_\text{J} E_\text{C}}$). As $f_\text{t}$ approaches the resonator the dispersive shift $\chi$ increases in magnitude, eventually switching sign as $f_\text{t}$ passes through $f_\text{r}$ (Fig \ref{leadalignment}a). In contrast, applying the field parallel to the film and along the long axis of the leads (see Fig 1c) the junction is able to retain a stable $E_\text{J}$ up to $B^{||_{\text{Contacts}}}_{||_{\text{Film}}} = \SI{1}{\tesla}$ (Fig 4b). 

\nocite{*}
\begin{thebibliography}{10}
	
	\bibitem{Koch2007}
	J.~Koch, T.~Yu, J.~Gambetta, A.~A. Houck, D.~I. Schuster, J.~Majer, A.~Blais,
	M.~H. Devoret, S.~M. Girvin, and R.~J. Schoelkopf.
	\newblock {Charge-insensitive qubit design derived from the Cooper pair box}.
	\newblock {\em Physical Review A}, 76(4):042319, 2007.
	
	\bibitem{Hassler2011}
	F.~Hassler, A.~R. Akhmerov, and C.~W.~J. Beenakker.
	\newblock {The top-transmon: a hybrid superconducting qubit for
		parity-protected quantum computation}.
	\newblock {\em New Journal of Physics}, 13(9):095004, 2011.
	
	\bibitem{Hyart2013}
	T.~Hyart, B.~{Van Heck}, I.~C. Fulga, M.~Burrello, A.~R. Akhmerov, and C.~W.~J.
	Beenakker.
	\newblock {Flux-controlled quantum computation with Majorana fermions}.
	\newblock {\em Physical Review B}, 88(3):035121, 2013.
	
	\bibitem{Aasen}
	D.~Aasen, M.~Hell, R.~Mishmash, A.~Higginbotham, J.~Danon, M.~Leijnse,
	T.~Jespersen, J.~Folk, C.~Marcus, K.~Flensberg, and J.~Alicea.
	\newblock {Milestones toward Majorana-based quantum computing}.
	\newblock {\em Physical Review X}, 6(3), 2016.
	
	\bibitem{Kubo2011}
	Y.~Kubo, C.~Grezes, A.~Dewes, T.~Umeda, J.~Isoya, H.~Sumiya, N.~Morishita,
	H.~Abe, S.~Onoda, T.~Ohshima, V.~Jacques, A.~Dr{\'{e}}au, J.~F. Roch,
	I.~Diniz, A.~Auffeves, D.~Vion, D.~Esteve, and P.~Bertet.
	\newblock {Hybrid quantum circuit with a superconducting qubit coupled to a
		spin ensemble}.
	\newblock {\em Physical Review Letters}, 107(22):220501, 2011.
	
	\bibitem{Ranjan2013}
	V.~Ranjan, G.~{De Lange}, R.~Schutjens, T.~Debelhoir, J.~P. Groen, D.~Szombati,
	D.~J. Thoen, T.~M. Klapwijk, R.~Hanson, and L.~Dicarlo.
	\newblock {Probing dynamics of an electron-spin ensemble via a superconducting
		resonator}.
	\newblock {\em Physical Review Letters}, 110(6):067004, 2013.
	
	\bibitem{Calado2015}
	V.~E. Calado, S.~Goswami, G.~Nanda, M.~Diez, A.~R. Akhmerov, K.~Watanabe,
	T.~Taniguchi, T.~M. Klapwijk, and L.~M.K. Vandersypen.
	\newblock {Ballistic Josephson junctions in edge-contacted graphene}.
	\newblock {\em Nature Nanotechnology}, 10(9):761--764, 2015.
	
	\bibitem{Allen2015}
	M.~T. Allen, O.~Shtanko, I.~C. Fulga, J.~I.~J. Wang, D.~Nurgaliev, K.~Watanabe,
	T.~Taniguchi, A.~R. Akhmerov, P.~Jarillo-Herrero, L.~S. Levitov, and
	A.~Yacoby.
	\newblock {Observation of electron coherence and Fabry-Perot standing waves at
		a graphene edge}.
	\newblock {\em Nano Letters}, 17(12):7380--7386, 2017.
	
	\bibitem{Chtchelkatchev2007}
	N.~M. Chtchelkatchev and I.~S. Burmistrov.
	\newblock {Conductance oscillations with magnetic field of a two-dimensional
		electron gas superconductor junction}.
	\newblock {\em Physical Review B}, 75(21):214510, 2007.
	
	\bibitem{Ku2016}
	J.~Ku, Z.~Yoscovits, A.~Levchenko, J.~Eckstein, and A.~Bezryadin.
	\newblock {Decoherence and radiation-free relaxation in Meissner transmon qubit
		coupled to Abrikosov vortices}.
	\newblock {\em Physical Review B}, 94(16):165128, 2016.
	
	\bibitem{Luthi2017}
	F.~Luthi, T.~Stavenga, O.~W. Enzing, A.~Bruno, C.~Dickel, N.~K. Langford, M.~A.
	Rol, T.~S. Jespersen, J.~Nyg{\aa}rd, P.~Krogstrup, and L.~DiCarlo.
	\newblock {Evolution of nanowire transmon qubits and their coherence in a
		magnetic field}.
	\newblock {\em Physical Review Letters}, 120(10), 2018.
	
	\bibitem{Lutchyn2010}
	R.~M. Lutchyn, J.~D. Sau, and S.~{Das Sarma}.
	\newblock {Majorana fermions and a topological phase transition in
		semiconductor-superconductor heterostructures}.
	\newblock {\em Physical Review Letters}, 105(7):077001, 2010.
	
	\bibitem{VanWoerkom2015}
	D.~J. {Van Woerkom}, A.~Geresdi, and L.~P. Kouwenhoven.
	\newblock {One minute parity lifetime of a NbTiN Cooper-pair transistor}.
	\newblock {\em Nature Physics}, 11(7):547--550, 2015.
	
	\bibitem{Kurizki}
	G.~Kurizki, P.~Bertet, Y.~Kubo, K.~M{\o}lmer, D.~Petrosyan, P.~Rabl, and
	J.~Schmiedmayer.
	\newblock {Quantum technologies with hybrid systems}.
	\newblock {\em Proceedings of the National Academy of Sciences},
	112(13):3866--3873, 2015.
	
	\bibitem{Riste2012}
	D.~Rist{\`{e}}, C.~C. Bultink, M.~J. Tiggelman, R.~N. Schouten, K.~W. Lehnert,
	and L.~Dicarlo.
	\newblock {Millisecond charge-parity fluctuations and induced decoherence in a
		superconducting transmon qubit}.
	\newblock {\em Nature Communications}, 4, 2013.
	
	\bibitem{Doh2005}
	Y.~Doh, J.~{Van Dam}, A.~Roest, E.~P. A.~M. Bakkers, L.~P. Kouwenhoven, and
	S.~{De Franceschi}.
	\newblock {Applied physics: Tunable supercurrent through semiconductor
		nanowires}.
	\newblock {\em Science}, 309(5732):272--275, 2005.
	
	\bibitem{VanWoerkom2017}
	D.~J. {Van Woerkom}, A.~Proutski, B.~{Van Heck}, D.~Bouman, J.~I.
	V{\"{a}}yrynen, L.~I. Glazman, P.~Krogstrup, J.~Nyg{\aa}rd, L.~P.
	Kouwenhoven, and A.~Geresdi.
	\newblock {Microwave spectroscopy of spinful Andreev bound states in ballistic
		semiconductor Josephson junctions}.
	\newblock {\em Nature Physics}, 13(9):876--881, 2017.
	
	\bibitem{Hays2017}
	M.~Hays, G.~de~Lange, K.~Serniak, D.~J. van Woerkom, D.~Bouman, P.~Krogstrup,
	J.~Nyg{\aa}rd, A.~Geresdi, and M.~H. Devoret.
	\newblock {Direct microwave measurement of Andreev-bound-state dynamics in a
		proximitized semiconducting nanowire}.
	\newblock 2017.
	
	\bibitem{delange2015}
	G.~{De Lange}, B.~{Van Heck}, A.~Bruno, D.~J. {Van Woerkom}, A.~Geresdi, S.~R.
	Plissard, E.~P.A.M. Bakkers, A.~R. Akhmerov, and L.~DiCarlo.
	\newblock {Realization of microwave quantum circuits using hybrid
		superconducting-semiconducting nanowire Josephson elements}.
	\newblock {\em Physical Review Letters}, 115(12):127002, 2015.
	
	\bibitem{larsen2015}
	T.~W. Larsen, K.~D. Petersson, F.~Kuemmeth, T.~S. Jespersen, P.~Krogstrup,
	J.~Nyg{\aa}rd, and C.~M. Marcus.
	\newblock {Semiconductor-nanowire-based superconducting qubit}.
	\newblock {\em Physical Review Letters}, 115(12):127001, 2015.
	
	\bibitem{Stan2004}
	G.~Stan, S.~B. Field, and J.~M. Martinis.
	\newblock {Critical field for complete vortex expulsion from narrow
		superconducting strips}.
	\newblock {\em Phys. Rev. Lett.}, 92:97003--97004, 2004.
	
	\bibitem{Bothner2011}
	D.~Bothner, T.~Gaber, M.~Kemmler, D.~Koelle, and R.~Kleiner.
	\newblock {Improving the performance of superconducting microwave resonators in
		magnetic fields}.
	\newblock {\em Applied Physics Letters}, 98(10):102504, 2011.
	
	\bibitem{Reed}
	M.~D. Reed, L.~Dicarlo, B.~R. Johnson, L.~Sun, D.~I. Schuster, L.~Frunzio, and
	R.~J. Schoelkopf.
	\newblock {High-fidelity readout in circuit quantum electrodynamics using the
		jaynes-cummings nonlinearity}.
	\newblock {\em Physical Review Letters}, 105(17), 2010.
	
	\bibitem{Nanda2017}
	G.~Nanda, J.~L. Aguilera-Servin, P.~Rakyta, A.~Korm{\'{a}}nyos, R.~Kleiner,
	D.~Koelle, K.~Watanabe, T.~Taniguchi, L.~M.K. Vandersypen, and S.~Goswami.
	\newblock {Current-phase relation of ballistic graphene Josephson junctions}.
	\newblock {\em Nano Letters}, 17(6):3396--3401, 2017.
	
	\bibitem{Kringhøj2018}
	A.~Kringh{\o}j, L.~Casparis, M.~Hell, T.~W. Larsen, F.~Kuemmeth, M.~Leijnse,
	K.~Flensberg, P.~Krogstrup, J.~Nyg{\aa}rd, K.~D. Petersson, and C.~M. Marcus.
	\newblock {Anharmonicity of a superconducting qubit with a few-mode Josephson
		junction}.
	\newblock {\em Physical Review B}, 97(6):060508, 2018.
	
\end{thebibliography}

\newpage

\begin{thebibliography}{1}
	
	\bibitem{Megrant2012}
	A~Megrant, C~Neill, R~Barends, B~Chiaro, Yu~Chen, L~Feigl, J~Kelly, Erik
	Lucero, Matteo Mariantoni, P~J~J {O 'malley}, D~Sank, A~Vainsencher,
	J~Wenner, T~C White, Y~Yin, J~Zhao, C~J Palmstr{\o}m, John~M Martinis, and
	A~N Cleland.
	\newblock {Planar Superconducting Resonators with Internal Quality Factors
		above One Million}.
	\newblock 2012.
	
	\bibitem{Singh2014}
	Vibhor Singh, Ben~H. Schneider, Sal~J. Bosman, Evert~P.J. Merkx, and Gary~A.
	Steele.
	\newblock {Molybdenum-rhenium alloy based high- Q superconducting microwave
		resonators}.
	\newblock {\em Applied Physics Letters}, 105(22):222601, dec 2014.
	
	\bibitem{Stan2004}
	Gheorghe Stan, Stuart~B. Field, and John~M. Martinis.
	\newblock {Critical Field for Complete Vortex Expulsion from Narrow
		Superconducting Strips}.
	\newblock {\em Physical Review Letters}, 92(9):097003, mar 2004.
	
	\bibitem{Samkharadze2016}
	N.~Samkharadze, A.~Bruno, P.~Scarlino, G.~Zheng, D. P. DiVincenzo,
	L.~DiCarlo, and L. M. K. Vandersypen.
	\newblock {High-Kinetic-Inductance Superconducting Nanowire Resonators for
		Circuit QED in a Magnetic Field}.
	\newblock {\em Physical Review Applied}, 5(4):044004, apr 2016.
	
	\bibitem{Bothner2011}
	D.~Bothner, T.~Gaber, M.~Kemmler, D.~Koelle, and R.~Kleiner.
	\newblock {Improving the performance of superconducting microwave resonators in
		magnetic fields}.
	\newblock {\em Applied Physics Letters}, 98(10):102504, mar 2011.
	
	\bibitem{Calado2015}
	V.~E. Calado, S.~Goswami, G.~Nanda, M.~Diez, A.~R. Akhmerov, K.~Watanabe,
	T.~Taniguchi, T.~M. Klapwijk, and L.~M.K. Vandersypen.
	\newblock {Ballistic Josephson junctions in edge-contacted graphene}.
	\newblock {\em Nature Nanotechnology}, 10(9):761--764, sep 2015.
	
\end{thebibliography}
\end{document}